\documentclass[preprint,showpacs]{revtex4}

\usepackage{graphicx}
\usepackage{amsmath}
\usepackage{dcolumn}
\usepackage{color}

\begin{document}

\title{{Tunable pulse delay and advancement in a coupled nanomechanical resonator-superconducting microwave cavity system}
}
\author{Cheng Jiang}
\author{Bin Chen}
\author{Ka-Di Zhu}
\email{zhukadi@sjtu.edu.cn}
\affiliation{Key Laboratory of
Artificial Structures and Quantum Control (MOE), Department of
Physics, Shanghai Jiao Tong University, 800 Dong Chuan Road,
Shanghai 200240, China }
\date{\today}

\begin{abstract}
We theoretically study the transmission of a weak probe field under
the influence of a strong pump field in a coupled nanomechanical
resonator-superconducting microwave cavity system. Using the
standard input-output theory, we find that both pulse delay (slow
light effect) and advancement (fast light effect) of the probe field
can appear in this coupled system provided that we choose the
suitable detuning of the pump field from cavity resonance. The
magnitude of the delay (advancement) can be tuned continuously by
adjusting the power of the pump field. This technique demonstrates
great potential in applications including microwave phase shifter
and delay line.
\end{abstract}

\maketitle

   Control of slow and fast light has received a lot of
interest in view of its importance in understanding the physical
laws that govern how fast or slow a light pulse can be made to
propagate and its potential impact on photonic technology
\cite{Kuzmich, Mok}. Various techniques have been developed to
realize slow light and fast light in atomic vapors and solid state
materials. Early studies on slow light have made use of the
technique of electromagnetically induced transparency (EIT) in
atomic vapors \cite{Kasapi} or Bose-Einstein condensate
(BEC)\cite{Hau}. And coherent population oscillation (CPO) is
another mechanism that makes slow and fast light possible. Using
this technique, Bigelow \textit{et al.} observed a group velocity of
58m/s in a ruby crystal \cite{Bigelow1} and they also observed both
superluminal and ultraslow light propagation in an alexandrite
crystal at room temperature \cite{Bigelow2}. Recently, stimulated
Brillouin scattering (SBS) was proposed as an efficient way to
realize slow and fast light due to its unique advantages such as
arbitrary choice of the a resonant frequency by changing the pump
frequency and room-temperature operation. Okawachi \textit{et
al.}\cite{Okawachi} demonstrated that SBS in a single-mode fiber
could be used to induce tunable all-optical slow-light pulse delay
as much as 25$ns$. Observations of both slow and fast light in
optical fibers using SBS were also reported \cite{Song,
Herraez,Thevenaz}.

    On the other hand, nanomechanical resonators (NRs) are currently
under intensive exploration owing to their combination of large
quality factors ($10^{3}\sim 10^{5}$) and high natural frequencies
(MHz $\sim $\ GHz) together with the important applications \cite{K.
C. Schwab,T. J. Kippenberg}, such as high precision displacement
detection \cite{C. M. Caves,M. F. Bocko}, zeptogram-scale mass
sensing \cite{Yang} and quantum measurements \cite{V. B. Braginsky}.
Quite recently, the analog of electromagnetically induced
transparency (EIT) in Fabry-Perot cavity \cite{Agarwal1} as well as
in whispering-gallery-mode (WGM) microresonators \cite{Schliesser,
Weis} has been studied. Alternatively, in the present paper, we
theoretically investigate the microwave pulse advancement and delay
based on the analog of EIT in the coupled nanomechanical resonator
(NR)-superconducting microwave cavity (SMC) system \cite{Regal}.
Much attention has been paid to this coupled system to investigate
quantum entanglement, nanomechanical squeezing and cooling of the
nanomechanical resonator \cite{Vitali,Woolley,Rocheleau}. Here, we
propose a scheme to efficiently switch from pulse delay to pulse
advancement by simply adjusting the pump-cavity detuning. Our
results indicate some potential applications in microwave photonics
including microwave phase shifter and delay line \cite{Xue,Wei}.

The system under consideration, sketched in figure 1, is a
nanomechanical resonator with resonance frequency $\omega_{n}$ and
effective mass $m$ capacitively coupled to a superconducting
microwave cavity denoted by the equivalent inductance $L$ and
equivalent capacitance $C$. A strong pump field with frequency
$\omega_{p}$ and a weak probe field with frequency $\omega_{r}$
drive the microwave cavity simultaneously. The experiment of such a
scheme is usually operated in a dilution refrigerator where the
thermomechanical motion is greatly reduced and the transmitted sinal
is probed through a low-noise high-electron-mobility-transistor
(HEMT) microwave amplifier with a homodyne detection scheme. The
displacement $x$ of the nanomechanical resonator from its
equilibrium position alters the capacitance of the microwave cavity
and therefore its resonance frequency. The coupling capacitance can
be approximated by $C_0(x)=C_0(1-x/d)$, where $C_{0}$\ represents an
equilibrium capacitance and \textit{d} is the equilibrium
nanoresonator-cavity separation, thus the coupled cavity has an
equivalent capacitance $C_\Sigma=C+C_0$, such that resonance
frequency $\omega_{c}=1/\sqrt{LC_\Sigma}$.

    In a rotating frame at the pump frequency $\omega_p$, the Hamiltonian
of the coupled system is given by \cite{Regal,Vitali}
{\setlength\arraycolsep{1pt}
\begin{eqnarray}
H=\hbar\Delta_pa^\dagger a+\hbar\omega_nb^\dagger b-\hbar\lambda
a^\dagger aQ+i\hbar(E_pa^\dagger-E_p^*a)+i\hbar(E_ra^\dagger
e^{-i\delta t}-E_r^*ae^{i\delta t}),
\end{eqnarray}
Here $a^\dagger$ $(a)$ and $b^\dagger$ $(b)$ are the creation
(annihilation) operators of the microwave cavity and nanomechanical
resonator, respectively. The first two terms describe the energy of
the microwave cavity and the nanomechanical resonator, where
$\Delta_p=\omega_c-\omega_p$ is the detuning of the microwave cavity
and the pump field. The third term corresponds to the capacitive
coupling between the microwave cavity and the nanomechanical
resonator, where $\lambda=g\delta x_{zp}$ is the coupling strength
between the cavity and the resonator
($g=\frac{\partial\omega_c}{\partial x}$ is the effect of the
displacement $x=(b^\dagger+b)\delta x_{zp}$ on the perturbed cavity
resonance frequency, $\delta x_{zp}=\sqrt{\frac{\hbar}{2\omega_nm}}$
is the zero-point motion of the nanomechanical resonator) and
$Q=b^\dagger+b$ is the phonon amplitude of the resonator. The last
two terms give the interaction of the cavity field with the pump
field and the probe field, $\delta=\omega_r-\omega_p$ is the
detuning of the probe and the pump field, $E_{p}$ and $E_{r}$ are,
respectively, amplitudes of the pump field and probe field. They are
defined by $\left\vert
E_p\right\vert=\sqrt{2P_p\kappa/\hbar\omega_p}$ and $\left\vert
E_r\right\vert=\sqrt{2P_r\kappa/\hbar\omega_r}$, where $P_{p}$ is
the pump power, $P_{r}$ is the probe power, $\kappa$ is the
linewidth of the microwave cavity.

   Let $\left\langle a\right\rangle$, $\left\langle
a^\dagger\right\rangle$, and $\left\langle Q\right\rangle$ be the
expectation values of the operators $a$, $a^\dagger$, and $Q$,
respectively. And in what follows we ignore the quantum properties
of $a$ and $Q$ \cite{B. I. Greene,J. F. Lam,G. S. Agarwal}, the time
evolutions of these expectation values can be obtained by employing
the Heisenberg equation of motion and by addition of the damping
terms phenomenologically,
\begin{eqnarray}
&&\frac{d\left\langle
a\right\rangle}{dt}=-\left(i\Delta_p+\kappa\right)\left\langle
a\right\rangle+i\lambda\left\langle a\right\rangle\left\langle
Q\right\rangle+E_p+E_re^{-i\delta t},\\
&&\frac{d^2\left\langle
Q\right\rangle}{dt^2}+\gamma_n\frac{d\left\langle
Q\right\rangle}{dt}+\omega_n^2\left\langle
Q\right\rangle=2\omega_n\lambda\left\langle
a^\dagger\right\rangle\left\langle a\right\rangle,
\end{eqnarray}
where $\gamma_n$ is the damping rate of the mechanical mode. In
order to solve Eq.(2) and Eq.(3), we make the ansatz \cite{R. W.
Boyd} $\langle a(t)\rangle=a_0+a_+e^{-i\delta t}+a_-e^{i\delta t}$,
and $\langle Q(t)\rangle=Q_0+Q_+e^{-i\delta t}+Q_-e^{i\delta t}$
with the relationship $\left\vert a_+\right\vert$, $\left\vert
a_-\right\vert$ $\ll$ $\left\vert a_0\right\vert$ and $\left\vert
Q_+\right\vert$, $\left\vert Q_-\right\vert$ $\ll$ $\left\vert
Q_0\right\vert$. Since the probe field $E_{r}$ is much weaker than
the pump field $E_{p}$, we derive the steady state solution of the
Eq.(2) and Eq.(3) upon substituting the above ansatz into it and
upon working to the lowest order in $E_r$ but to all orders in
$E_p$,
\begin{eqnarray}
a_+=\frac{\delta+\Delta_p+i(\kappa+\theta)}{(\delta+i\kappa)^2+(\theta-i\Delta_p)^2+\beta}i
E_r,
\end{eqnarray}
where $\eta=\frac{\omega_n^2}{\omega_n^2-\delta^2-i\gamma_n\delta}$, $\alpha=\frac{2\lambda^2}{\omega_n^2}$,
$\beta=\alpha^2\eta^2\omega_n^2n_p^2$, $\theta=i\alpha\omega_nn_p(\eta+1)$ and $n_p=\left\vert
a_0\right\vert^2$, approximately equal to the cavity photon
occupation, is determined by the equation
\begin{eqnarray}
n_p\left[\kappa^2+\left(\Delta_p-\omega_{n}\alpha
n_p\right)^2\right]=\left\vert E_p\right\vert^2.
\end{eqnarray}

Using the standard input-output theory \cite{Gardiner}
$a_{out}(t)=a_{in}(t)-\sqrt{2\kappa}a(t)$, where $a_{out}(t)$ is the
output field operator, we obtain
\begin{eqnarray}
\left\langle a_{out}(t)\right\rangle&=&(E_p-\sqrt{2\kappa}a_0)
e^{-i\omega t}+(E_r-\sqrt{2\kappa}a_+)
e^{-(\delta+\omega_p)t}-\sqrt{2\kappa}a_- e^{i(\delta-\omega_p)t}.
\end{eqnarray}
The transmission of the probe field, defined by the ratio of the
output and input field amplitudes at the probe frequency, is then
given by
\begin{eqnarray}
t_{p}=\frac{E_r-\sqrt{2\kappa}a_+}{E_r}=1-\sqrt{2\kappa}\frac{i(\delta+\Delta_p)-(\kappa+\theta)}{(\delta+i\kappa)^2+(\theta-i\Delta_p)^2+\beta}.
\end{eqnarray}
The tunable probe transmission window will modify the propagation
dynamics of a probe pulse sent to this coupled system due to the
variation of the complex phase picked by its different frequency
components. The probe pulse will experiences a group delay
$\tau_{g}$, and this group delay $\tau_{g}$ is defined by
$\tau_{g}=\left.\frac{d\phi}{d\omega_{r}}\right|_{\omega_{c}}$,
where $\phi(\omega_{r})$=arg($t_{p}(\omega_{r})$) is the rapid phase
dispersion. The magnitude and phase of the transmitted probe signal
could be determined experimentally by measuring the in-phase and
quadrature response of the system to the input modulation.

In order to demonstrate our numerical results, we choose a realistic
coupled nanomechanical resonator-superconducting microwave cavity
system with the parameters as follows \cite{Rocheleau}:
$\omega_{c}=2\pi\times7.5$ GHz, $\omega_{n}=2\pi\times6.3$ MHz,
$\kappa=2\pi\times600$ kHz, $\lambda=250$ Hz, and $Q_{n}=10^{6},$
where $Q_{n}$ is the quality factor of the nanomechanical resonator,
and the damping rate $\gamma _{n}$ is given by $\frac{\omega
_{n}}{Q_n}.$ The system therefore operates in the resolved-sideband
regime ($\omega_{n}>\kappa$) also termed good-cavity limit, which is
a prerequisite for cooling of the nanomechanical resonator.

   First, we consider the situation where the cavity is driven on its red sideband, i.e.,
$\Delta_{p}=\omega_{n}$, leading to up-conversion of the pump
photons to $\omega_{c}.$ Figure 2(a) shows the transmission and
phase dispersion of the transmitted probe field as a function of
probe-cavity detuning $\Delta_{r}=\omega_{r}-\omega_{c}$ for pump
power $P_{p}=8nW$ and $\Delta_{p}=\omega_{n}$. We find that there is
a narrow transparency window with a very steep positive phase
dispersion around $\Delta_{r}=0$. As a result, there should be slow
light effect in this region. This phenomenon is a result of the
mechanical analogy of electromagnetically induced transparency
(EIT). The simultaneous presence of pump and probe fields generates
a radiation force at the beat frequency $\delta$ resonant with the
mechanical resonant frequency $\omega_n$. The frequency of the pump
field $\omega_{p}$ is upshifted to the anti-stokes frequency
$\omega_{p}+\omega_{n}$, which is degenerate with the probe field.
Destructive interference between the anti-stokes field and the probe
field can suppress the build-up of an intracavity probe field and
result in the narrow transparency window. Here, $\delta=\omega_n$
satisfies the two-photon resonance condition. Furthermore, the
effective coupling strength between the microwave cavity and
nanoresonator will enlarge with larger number of cavity photon
occupation when the pump power increases, and the transparency
window will become broader. On the other hand, the displacement $x$
of the nanomechanical resonator from its equilibrium position alters
the capacitance of the microwave cavity and therefore its resonance
frequency. As a consequence, the effective refractive index seen by
the propagating probe field changes and a phase shift is induced. In
figure 2(b), we plot the group delay $\tau_{g}$ as a function of the
pump power. As the pump power increases, the group delay decreases,
and we can get the group delay with the magnitude as much as 0.2
$ms$ at a very low pump power. Therefore, we can tune the probe
pulse delay by adjusting the pump power if we set
$\Delta_{p}=\omega_{n}$.

   Furthermore, if the cavity is driven on its blue sideband, i.e.,
$\Delta_{p}=-\omega_{n}$, leading to the down-conversion of the pump
photons to $\omega_{c}$. Similarly, we plot the transmission and
phase of the transmitted probe field as a function of the
probe-cavity detuning for $P_{p}=4nW$ and $\Delta_{p}=-\omega_{n}$
in figure 3(a). We find that there is also a narrow transparency
window but with a steep negative phase dispersion around
$\Delta_{r}=0$, which may result in fast light effect. Here, we have
$\delta=-\omega_{n}$, the frequency of the pump field is downshifted
to the stokes frequency $\omega_p-\omega_n$. Destructive
interference between the probe field and the stokes field leads to
the transparency window. Figure 3(b) shows the group delay
$\tau_{g}$ as a function of the pump power for
$\delta=\Delta_{p}=-\omega_{n}$. The group delay is negative,
therefore, we can obtain fast light effect when the cavity is driven
on its blue sideband. Time-advanced signals can be used to
compensate time delays inevitable in an complex optical-processing
network \cite{Solli}. It's worth noticing that the phenomenon of
pulse advancement can be counterintuitive owing to the presence of
phenomena where the peak of the output pulse will appear earlier
than the peak of the input pulse, its consistency with the principle
has been verified experimentally \cite{Stenner}. From figure 2 and
figure 3, one can fix the probe field with frequency
$\omega_{r}=\omega_{c}$, and then scan the pump frequency across the
cavity resonance frequency $\omega_{c}$, one can efficiently switch
from probe pulse delay to advancement without appreciable absorption
or amplification as the pump detuning $\Delta_{p}$ equals to
$\omega_{n}$ or $-\omega_{n}$.

In conclusion, we have investigated the tunable microwave pulse
delay and advancement in a coupled nanomechanical
resonator-superconducting microwave cavity system. When the cavity
is driven by a strong pump field on its red sideband or blue
sideband, the transmitted weak probe field will experience pulse
group delay or advancement. Furthermore, the magnitude of the delay
or advancement can be tuned by adjusting the power of the pump field
in a wide range. \vskip 2pc The authors gratefully acknowledge
support from National Natural Science Foundation of China
(No.10774101 and No.10974133) and the National Ministry of Education
Program for Training Ph.D.

\newpage \centerline{\large{\bf Figure Captions}}

Figure 1 (a) Schematic of a nanomechanical resonator capacitively
coupled to a microwave cavity in the form of a superconducting
coplanar waveguide(denoted by \textit{LC} circuit) in the presence
of a strong pump field $\omega_{p}$ and a weak probe field
$\omega_{r}$. (b) Equivalent circuit.

Figure 2 (a) The normalized magnitude and phase of the cavity
transmission as a function of probe-cavity detuning
$\Delta_{r}=\omega_{r}-\omega_{c}$ for pump power $P_{p}=8n$W and
$\Delta_{p}=\omega_{n}$. (b) Group delay as a function of the pump
power for $\delta=\Delta_{p}=\omega_{n}$. Other parameters are
$\omega_{c}=2\pi\times7.5$ GHz, $\kappa =2\pi\times 600$ kHz,
$\lambda=250$ Hz, $\gamma_{n}$=40 Hz and $\omega_{n}=2\pi\times6.3$
MHz.

Figure 3 (a) The normalized magnitude and phase of the cavity
transmission as a function of probe-cavity detuning
$\Delta_{r}=\omega_{r}-\omega_{c}$ for pump power $P_{p}=4n$W and
$\Delta_{p}=-\omega_{n}$. (b) Group delay as a function of the pump
power for $\delta=\Delta_{p}=-\omega_{n}$. Other parameters are
$\omega_{c}=2\pi\times7.5$ GHz, $\kappa =2\pi\times 600$ kHz,
$\lambda=250$ Hz, $\gamma_{n}$=40 Hz and $\omega_{n}=2\pi\times6.3$
MHz.

\clearpage
\begin{figure}[tbp]
\centerline{\includegraphics[width=14cm]{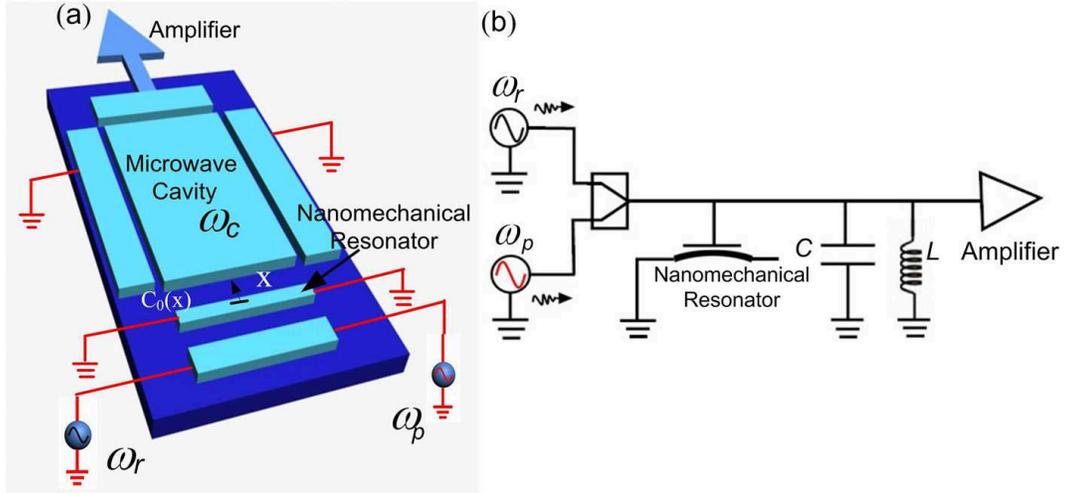}} \caption{(a)
Schematic of a nanomechanical resonator capacitively coupled to a
microwave cavity in the form of a superconducting coplanar
waveguide(denoted by \textit{LC} circuit) in the presence of a
strong pump field $\omega_{p}$ and a weak probe field $\omega_{r}$.
(b) Equivalent circuit.}
\end{figure}

\clearpage
\begin{figure}[tbp]
\centerline{\includegraphics[width=14cm]{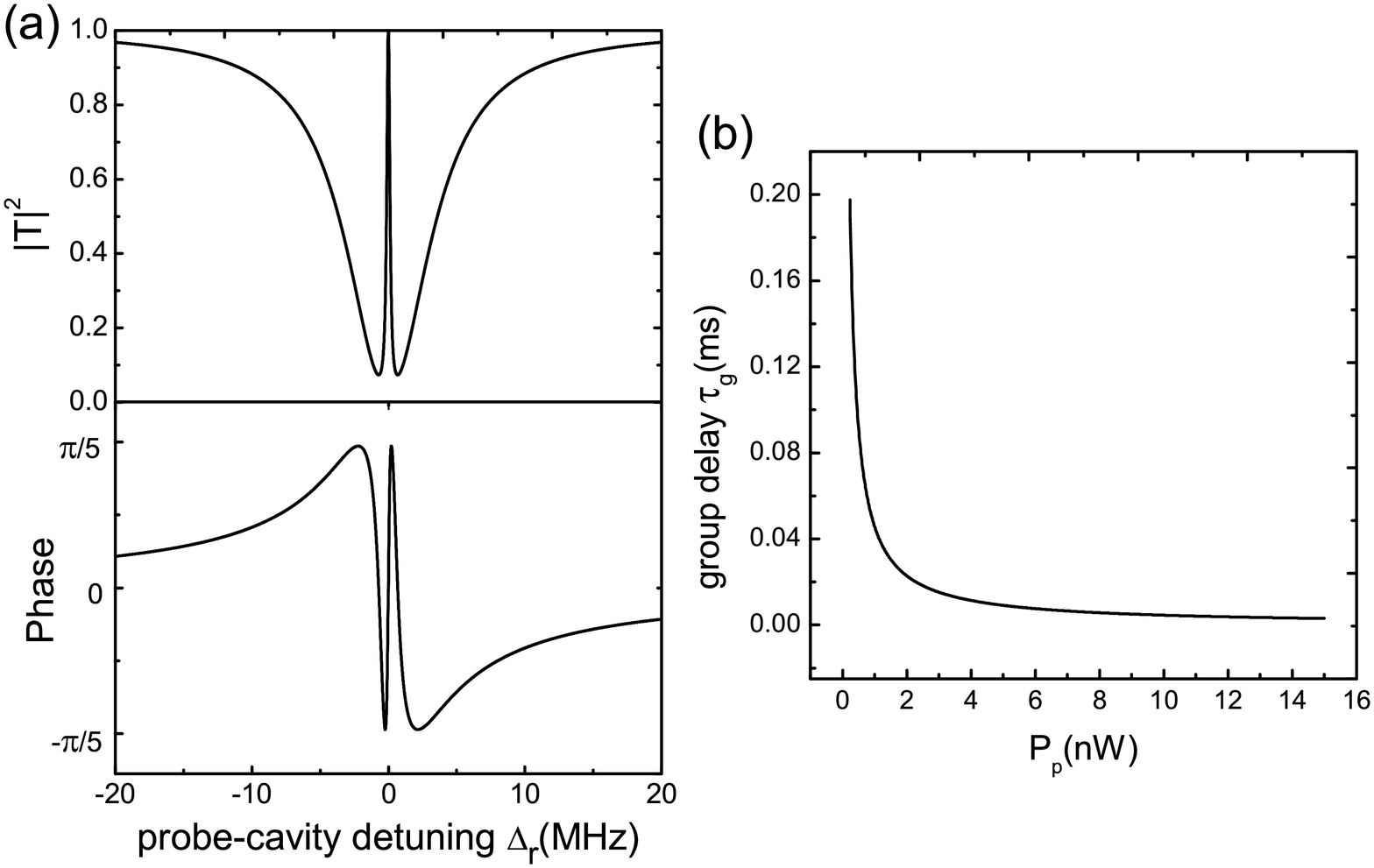}} \caption{(a)
The normalized magnitude and phase of the cavity transmission as a
function of probe-cavity detuning $\Delta_{r}=\omega_{r}-\omega_{c}$
for pump power $P_{p}=8n$W and $\Delta_{p}=\omega_{n}$. (b) Group
delay as a function of the pump power for
$\delta=\Delta_{p}=\omega_{n}$. Other parameters are
$\omega_{c}=2\pi\times7.5$ GHz, $\kappa =2\pi\times 600$ kHz,
$\lambda=250$ Hz, $\gamma_{n}$=40 Hz and $\omega_{n}=2\pi\times6.3$
MHz.}
\end{figure}

\clearpage
\begin{figure}[tbp]
\centerline{\includegraphics[width=14cm]{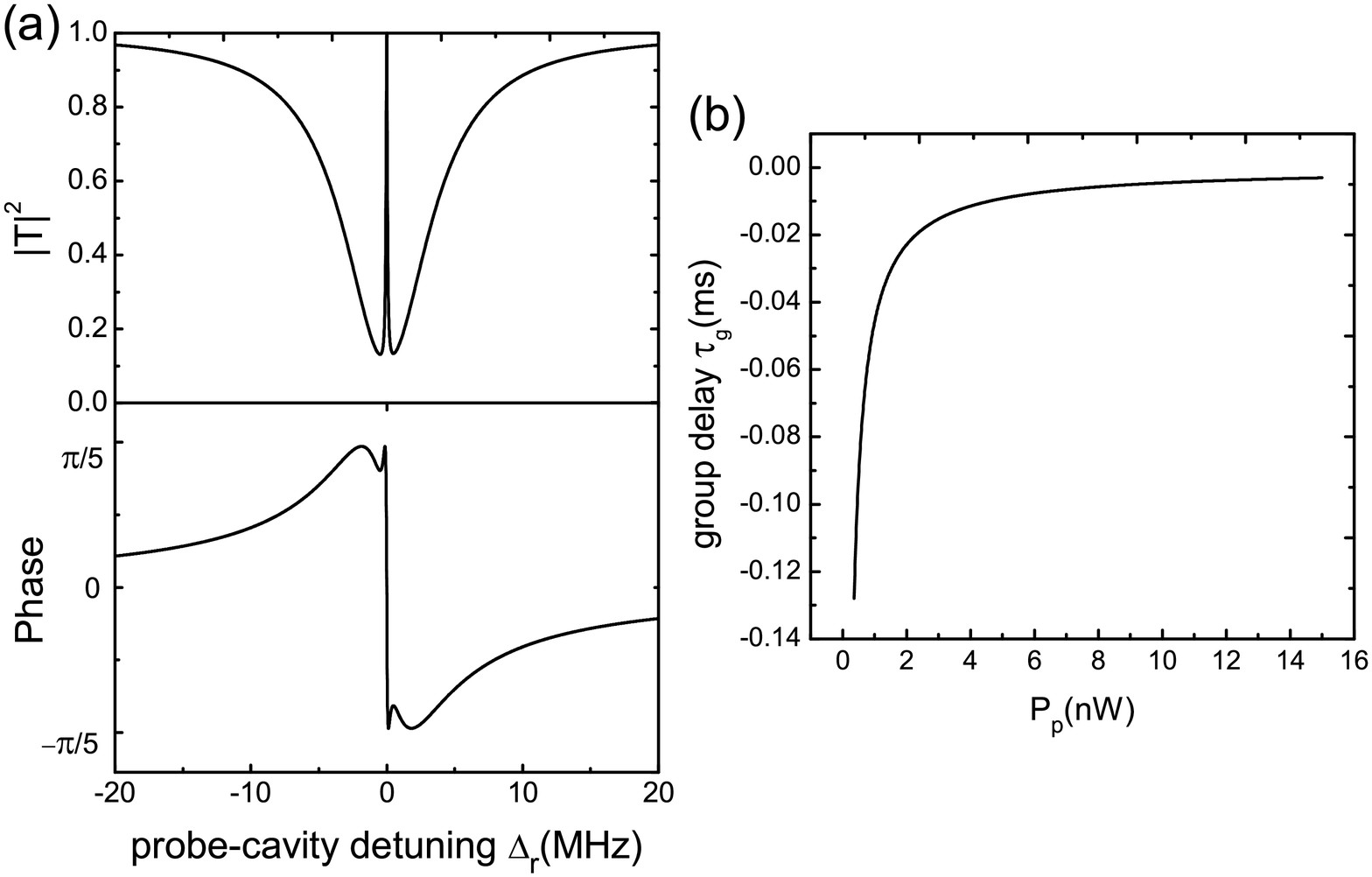}} \caption{(a)
The normalized magnitude and phase of the cavity transmission as a
function of probe-cavity detuning $\Delta_{r}=\omega_{r}-\omega_{c}$
for pump power $P_{p}=4n$W and $\Delta_{p}=-\omega_{n}$. (b) Group
delay as a function of the pump power for
$\delta=\Delta_{p}=-\omega_{n}$. Other parameters are
$\omega_{c}=2\pi\times7.5$ GHz, $\kappa =2\pi\times 600$ kHz,
$\lambda=250$ Hz, $\gamma_{n}$=40 Hz and $\omega_{n}=2\pi\times6.3$
MHz.}
\end{figure}

\end{document}